\documentclass{article}

\usepackage[french]{babel}

\usepackage[letterpaper,top=2cm,bottom=2cm,left=3cm,right=3cm,marginparwidth=1.75cm]{geometry}

\usepackage{amsmath}
\usepackage{amsfonts}
\usepackage{graphicx}
\usepackage{dsfont}
\usepackage{array}
\usepackage[colorlinks=true, allcolors=blue]{hyperref}
\usepackage[T1]{fontenc} 
\usepackage{mathtools}
\usepackage{authblk}

\usepackage{tikz}
\usetikzlibrary{shapes.geometric, arrows}
\tikzstyle{object} = [rectangle, rounded corners, minimum width=3cm, minimum height=1cm,text centered, draw=black, fill=red!30]
\tikzstyle{operator} = [trapezium, trapezium left angle=70, trapezium right angle=110, minimum width=3cm, minimum height=1cm, text centered, draw=black, fill=blue!30]
\tikzstyle{process} = [rectangle, minimum width=3cm, minimum height=1cm, text centered, draw=black, fill=orange!30]
\tikzstyle{decision} = [diamond, minimum width=3cm, minimum height=1cm, text centered, draw=black, fill=green!30]
\tikzstyle{arrow} = [thick,->,>=stealth]

\title{Enhancing Fluorescence Correlation Spectroscopy with Machine Learning for Advanced Analysis of Anomalous Diffusion}%Classifying and characterizing random walks from Fluorescence Correlation Spectroscopy with machine learning}

\author[1]{Nathan Quiblier}
\author[1]{Jan-Michael Rye}
\author[2]{Pierre Leclerc}
\author[2]{Henri Truong}
\author[2]{Abdelkrim Hannou}
\author[2]{Laurent Héliot}
\author[1,*]{Hugues Berry}
\affil[1]{AIstroSight, Inria, Hospices Civils de Lyon, Université Claude Bernard Lyon 1, F-69603 Villeurbanne, France}
\affil[2]{Univ. Lille, CNRS, UMR 8523, PhLAM Laboratoire de Physique des Lasers, Atomes et Molécules, F-59658, Lille, France}
\affil[*]{hugues.berry@inria.fr}

\begin{document}
\maketitle
\section{Abstract}
The random motion of molecules in living cells has consistently been reported to deviate from standard Brownian motion, a behavior coined as ``anomalous diffusion''. Fluorescence Correlation Spectroscopy (FCS) is a powerful method to quantify molecular motions in living cells but its application is limited to a subset of random motions and to long acquisition times. Here, we propose a new analysis approach that frees FCS of these limitations by using machine learning to infer the underlying model of motion and estimate the motion parameters. Using simulated FCS recordings, we show that this approach enlarges the range of anomalous motions available in FCS. We further validate our approach via experimental FCS recordings of calibrated fluorescent beads in increasing concentrations of glycerol in water. Taken together, our approach significantly augments the analysis power of FCS to capacities that are similar to the best-in-class state-of-the-art algorithms for single-particle-tracking experiments. 

\section{Introduction}
\label{intro}

Deviation of random motion from standard Brownian motion (BM) has received considerable attention in the literature to describe diverse physical situations \cite{hosking1984modeling,mandelbrot1968fractional,METZLER20001}. For instance, anomalous diffusion, where the mean-squared displacement scales non-linearly with time, $\left\langle r^2(t) \right\rangle=D t^\alpha$, has been reported to describe the motion of several proteins or particles in living cells \cite{Jeon2011,Izzedin,DiegoKrapf2015,Woringer2018,Sabri2020}. In this case, the exponent $\alpha$ is usually referred to as the anomalous exponent, and $D$ is the diffusion coefficient.  All anomalous subdiffusion motion models exhibit $\alpha<1$, whereas $\alpha=1$ for standard Brownian motion. However, anomalous subdiffusion is a characteristic shared by several unrelated types of motion. For instance, continuous-time random walk (CTRW), fractional Brownian motion (fBM) or random walk on a fractal support (RWf), all exhibit anomalous subdiffusion while the physical processes they describe are very different: heavy-tailed residence time distribution for CTRW, correlation between successive jumps for fBM or the fractal geometry of the object on which RWf takes place \cite{hofling2013anomalous,Woringer2020}. Therefore, the complete characterization of the motion of a biomolecule in a live cell requires the completion of two tasks: (\textit{i}) a classification or selection task to decide what model is the best at explaining the observations (e.g., BM, fBM, RWf or CTRW) and (\textit{ii}) an inference or calibration task, to estimate the parameter values of the selected model given an experimental observation.

In recent years, the advent of single-particle tracking supra-resolution microscopy~\cite{manzo2015review,shen2017single} has generalized the use of individual trajectories to quantify the motion of biomolecules or particles in living cells. A range of methods have been proposed for the classification and inference tasks based on individual trajectories \cite{cheng2022review}, from simple (non-)linear regression \cite{Izzedin,Fournier2023}, statistical tests \cite{Briane2018,Weron2019} or Bayesian inference \cite{Krog2018,Tharpa2018}, to machine- \cite{Janczura2020,LochOlszewska2020} and deep-learning \cite{Granik2019,Han2020}. A key factor here is the length of the observed individual trajectories, since for all the methods, the longer the individual trajectories, the better the performance. Experimentally, though, technical limits strongly constraint the typical time of  a trajectory, which can be as large as several seconds for membrane proteins \cite{Akin2016, Weron2017} but is usually closer to milliseconds for motions probed in the nucleus \cite{Izzedin,Fournier2023}. 

%The performances of state-of-the-art methods for single-particle tracking characterization have recently been benchmarked via an open community competition \cite{Nature}. Based on synthetic data in 1d, 2d or 3d, this open challenge featured three tasks: model classification (among 5 possible anomalous diffusion models), inference of the anomalous diffusion exponent and trajectory segmentation (when the model class changes along the trajectory). For the classification task, the best-in-class algorithms exhibited good performances, with $F_1$ scores between 0.6 (short trajectories of $L=40$ positions) and 0.9 (lengths $L>500$). Likewise, for the inference task, the best methods provided estimates for $\alpha$ with mean absolute errors ranging from 0.35 ($L=40$) down to $\approx 0.10$ ($L>500$), for values of $\alpha \in [0.05,2]$ (acquisition frame rate of $100$ frames per second).

On the other hand, Fluorescence Correlation Spectroscopy (FCS) is the main methodological alternative to single-particle-based techniques for the motion characterization of biomolecules in living cells \cite{krichevsky2002fluorescence,elson2011fluorescence}. In FCS, the biomolecules of interest are labelled with a fluorophore, and one monitors the fluctuations of the fluorescence signal due to their interaction with the light beam illuminating the sample. Although alternative approaches have been proposed \cite{Jiang2020}, data analysis in FCS is usually based on the auto-correlation of the fluorescence signal, $G(\tau)$. In the case of BM and fBM, theoretical considerations yield explicit non-linear functions for the expression of $G(\tau)$ as a function of the correlation delay $\tau$, the parameters of the optical setup and the parameters of the model of motion \cite{krichevsky2002fluorescence,hofling2013anomalous}. Fitting this expression to the measured auto-correlation can be used for both model classification and selection with information criteria as well as for parameter inference \cite{Weiss2003,Fournier2023}. 

Each approach, whether FCS or SPT, comes with its own specificity \cite{Woringer2020}. FCS can yield good results with a few individual molecules in the illumination volume, but is not a single-molecule approach, as opposed to SPT. The time scales they address are usually different: typically between 1 $\mu$s to 1 ms for FCS \textit{vs} 100 ms to 1 s for SPT. In SPT, one usually has to reconstruct the trajectories from the measured individual localizations. Tracking errors during these reconstructions can induce significant measurement errors \cite{rose2020particle}. In FCS, the signal-to-noise ratio of the auto-correlation function is usually low, so one has to continuously monitor the signal over long durations (more than 1 second) and average large numbers of consecutive measurements (often more than 100). Because of this, FCS is usually not able to track changes of the motion parameters if they occur over a time scale shorter than several minutes. Finally, analytical expressions for the auto-correlation function $G(\tau)$ are available for BM and fBM, but they are still lacking for other anomalous models, e.g. RWf or CTRW \cite{hofling2013anomalous}\footnote{Actually, an analytical expression can be obtained for motions defined by stationary processes with anomalous diffusion at all times and Gaussian distribution of the spatial displacements~\cite{hofling2013anomalous}. In practice, this usually restricts to fBM.}. Therefore, FCS is usually considered not to be applicable to the characterization of RWf or CTRW. 

Here we show that most of the above shortcomings of FCS for the classification and characterization of biomolecule motions can be overcome. Instead of fitting 
 the auto-correlation function by a theoretical expression, we use machine learning based on the auto-correlation function to perform the classification and inference tasks. With synthetic FCS data, we show that this approach renders FCS a powerful tool to distinguish between a range of standard and anomalous motions (BM, fBM and CTRW). The performance of our approach for the classification task and for parameter inference is found to be similar to the best-in-class state-of-the-art SPT algorithms on long trajectories. Our approach accommodates a wide range of FCS experimental setup parameters (beam width and illumination intensity) and uses recordings that are both unique (one recording per estimation) and short ($\geq$ 100-200 ms). We show that it can be used to accurately track changes of the parameter motions even with 1 Hz parameter-change frequency. Finally we apply the method on experimental data using calibrated beads in water with an increasing concentration of glycerol. Our predictions regarding the model of motion and physical parameters follow the Stokes-Einstein law and serves as a validation of our method.

\section{Results}

\subsection{Motion classification and parameter inference on synthetic data}
    We generated a learning set of more than 2.5 millions simulated FCS experiments, corresponding to 945 values of motion parameters $\alpha$ and $D$ sampled uniformly in $(0,1)$ and $(0,10]$, respectively (see sec.~\ref{mm:algo} for details on the generation of synthetic FCS data). For each pair of sampled parameters, 3 sets of trajectories were simulated with the following models: Brownian motion (BM, for which $\alpha$ was set to 1), fractional Brownian motion (fBM) and continuous-time random walk (CTRW) (more details on the models in sec.~\ref{mm:models}). One constant concern in this study was to develop a method that is robust enough to accommodate a wide range of experimental setup parameters, as encountered in the FCS laboratories worldwide. To this end, for each of the $945\times 3$ trajectories generated with the sampled parameters, we generated 900 FCS recordings by covering a wide range of experimental setup parameters: illumination beam waists $\omega_{xy} \in \{200, 225,250,275,300\}$ nm, and $\omega_z \in \{400,500,600\}$ nm and recording durations $T_\mathrm{obs} \in \{0.1, 0.25, 0.5, 0.75, 1,1.25,1.5,2\}$ s.

\begin{figure}
    		\center
   		 \includegraphics[scale=1]{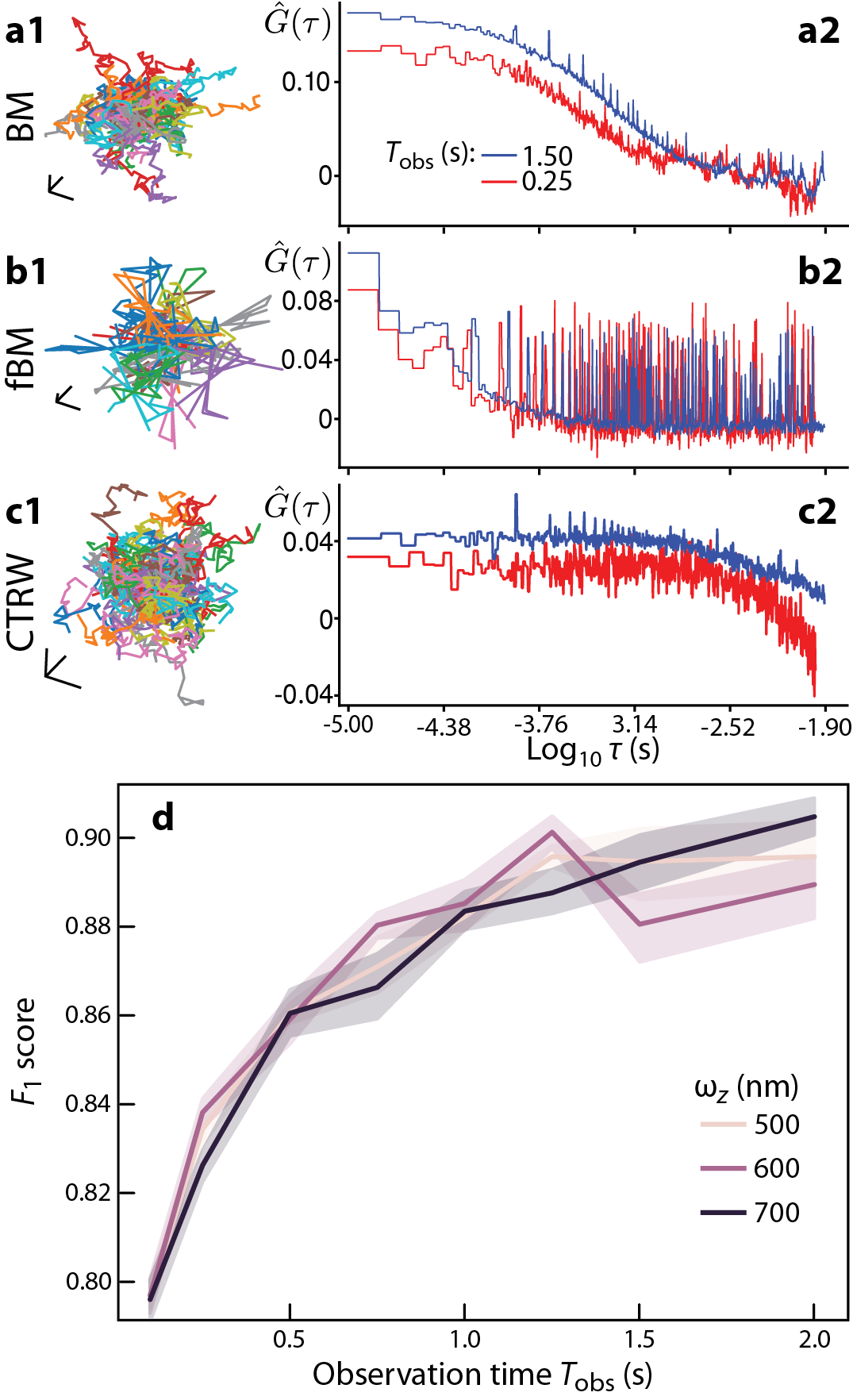}
   		 \caption{{\small Classification of motion models by FCS on synthetic data. (\textbf{a-c}) Illustrations of trajectories and corresponding features $\hat{G}(\tau)$ for Brownian motion (BM,~\textit{a}), fractional Brownian motion (fBM,~\textit{b}) and continuous-time random walks (CTRW,~\textit{c}). A classifier was trained to predict the model of motion for such synthetic data. Its performance as a function of  the duration of the FCS recordings $T_\mathrm{obs}$ is shown in (\textbf{d}), where full curves show the $F_1$ scores averaged over the test set, grouped by beam waist values $\omega_z$ as indicated in the legend. Shaded areas locate $\pm 1$ standard-deviation.  For the illustrations of (\textit{a1,\,b1,\,c1}), 50 trajectories of 50 steps were selected at random in the learning set, and their initial location set to $(0,0,0)$ for readability. Parameter values were $(D,\alpha)=(9.7,1)$ (\textit{a}), $(9.9,0.27)$ (\textit{b}) and $(7.3,0.57)$ (\textit{c}). In each panel the black lines represent (in each dimension): 50 (\textit{a1}), 500 (\textit{b1}) or 20 (\textit{c1}) nm. In (\textit{a2,\,b2,\,c2}), the features $\hat{G}(\tau)$ are shown for recordings of duration 0.25 (\textit{red}) or 1.50 (\textit{blue}) seconds.}}
   		 \label{fig:F1 score} 
    \end{figure}
    
    Figure~\ref{fig:F1 score} provides illustrations of the types of trajectories generated (Fig.~\ref{fig:F1 score}a1,b1,c1), as well as the corresponding estimators of the auto-correlation $\hat{G}(\tau)$ (defined by eq.~\ref{eq:Gest}) for simulated FCS recordings of 0.15 or 1.5 seconds (Fig.~\ref{fig:F1 score}a2,b2,c2). Due to the FCS signal-to-noise ratio (SNR) of the auto-correlation $\hat{G}(\tau)$, in fitting FCS methodologies, one typically accumulates and averages a large number of measurements (several hundreds), in order to, precisely, compensate for the lower SNR of individual measurements. Here, our objective was to test whether machine learning could exploit the information contained in individual auto-correlation measurements, despite their low SNR, in the absence of any averaging or accumulation procedure. 

    Figure~\ref{fig:F1 score}d shows the performance of our machine learning strategy for the model classification task. Our strategy, described in sec.~\ref{mm:ML_methods} is based on histogram gradient boosting and exclusively uses individual auto-correlation measurements as illustrated in Fig.~\ref{fig:F1 score}a2,b2,c2. Despite the low SNR of individual FCS recordings, our method exhibits very good classification accuracy, as measured by the $F_1$-score ($F_1=TP/[TP+0.5(FN+FP)]$, with $TP$ = \# true positives, $FN$= \# false negatives, $FP$=\# false positives). With observation times larger than 1.0 s, the average $F_1$-scores reach large values, in the range $[0.88-0.90]$. As expected, performance decreases with the FCS measurement time, but even with the smallest value used, $T_\mathrm{obs}=0.1$ s, the $F_1$-scores remain large, with values close to 0.80. Importantly, our algorithm manages to exhibit very similar values for all the beam waists tested, thus suggesting its applicability to a range of experimental setups. Indeed, the $F_1$-scores for the three $\omega_z$ of the figure are very similar. The shadings of these curves show the standard deviations of the score computed for different values of the motion parameters but also for different values of $\omega_{xy}$, the beam waist in the $x$ and $y$ direction. The amplitudes of the shadings reflect the fact that our algorithm delivers good performance for all the motion parameters and all the beam waists tested. The large values of the $F_1$-scores exhibited by our algorithm thus reveal its capacity to perform a robust classification of the motion types, even with individual (non-averaged) and short FCS measurements, and even when CTRW is part of the possible motions.

    \begin{figure}
    \center
   		 \includegraphics[scale=1]{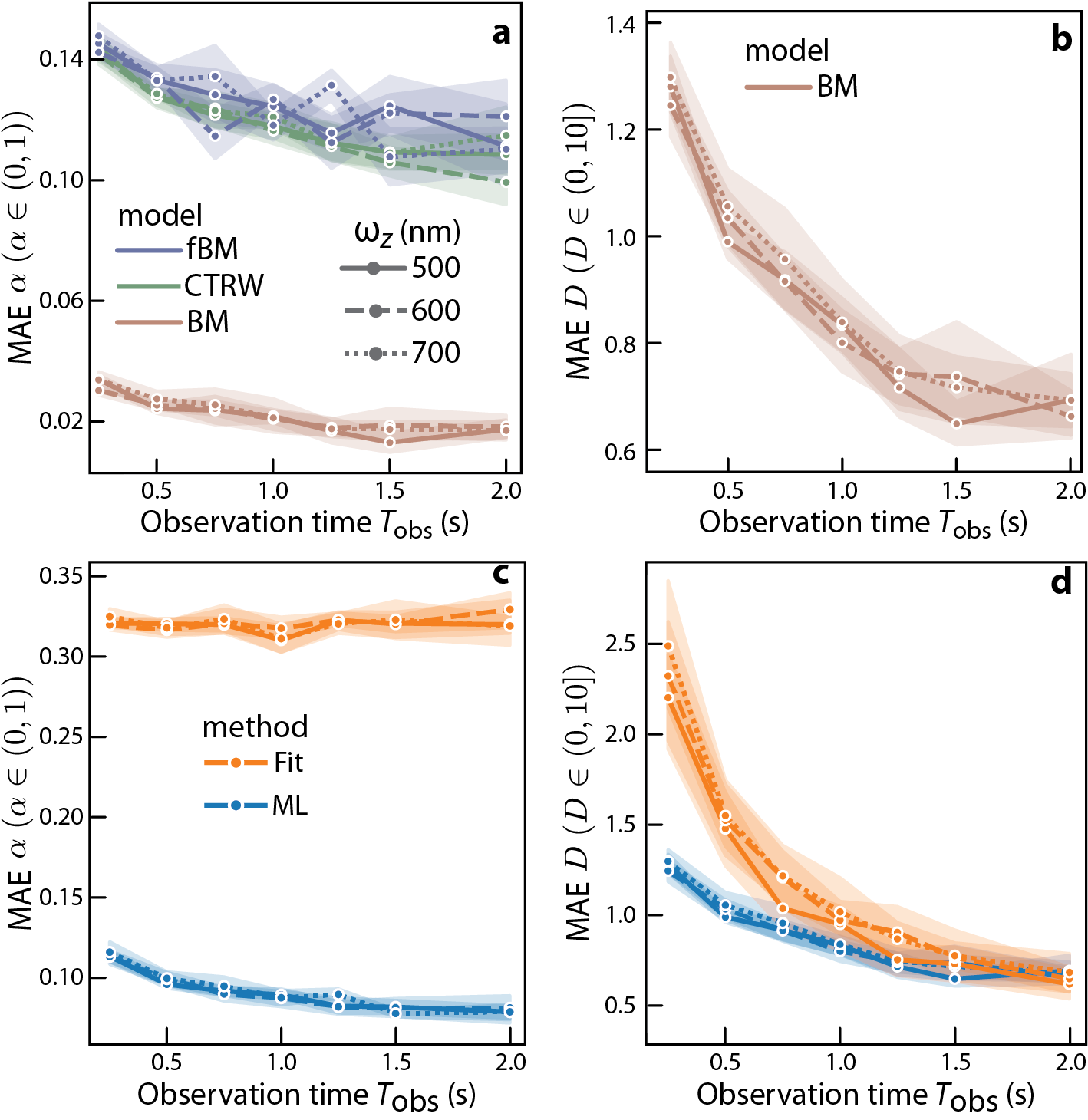}
   		 \caption{{\small Parameter estimation by FCS on synthetic data. A regressor was trained to estimate the motion parameters of the trajectories: the anomalous exponent $\alpha$ (\textbf{a}) and the diffusion coefficient $D$ (\textbf{b}). Its accuracy was computed as the mean absolute error (MAE), as a function of the duration of the FCS recordings $T_\mathrm{obs}$. Full curves show averages over the test set, grouped by beam waist values $\omega_z$ as indicated in the legend. Shaded areas locate $\pm 1$ standard-deviation. Comparison of the accuracy of the classical FCS method (non-linear fit, \textit{blue}) with our machine learning algorithm (ML, \textit{orange}) on synthetic data is shown in (\textbf{d}) for the anomalous exponent $\alpha$ and (\textbf{d}) for the diffusion coefficient $D$. In (a), the MAE is computed for all the data of the test set, using a colorcode that distinguishes the trajectories generated with fBM (brown), CTRW (indigo) or BM (green), independently of their classification, whereas panel (c) shows averages over all motion types (BM,fBM or CTRW) and their parameters. Only motions classified as BM are considered in (b) and (d).}}
   		 \label{fig:mae} 
    \end{figure}
    
   Regarding now the regression task, the accuracy of our machine learning algorithm is shown on Figure~\ref{fig:mae}, with separate inference of the anomalous exponent $\alpha$ (Fig.~\ref{fig:mae}a) and the diffusion coefficient (Fig.~\ref{fig:mae}b). The estimation of $\alpha$ exhibits very good accuracy with MAE (mean absolute error) values around 0.12 for the largest observation times, both for fBM and CTRW. Most of the BM trajectories are correctly classified (see  Fig~\ref{fig:F1 score}d), corresponding to $\alpha$ set to exactly 1, thus $MAE=0$. However, for the small fraction of BM trajectories that are incorrectly classified as fBM or CTRW, the inference of $\alpha$ yields values that are different but very close to $\alpha=1.0$. On average, the MAE for BM is therefore non-zero but still very small. In all cases, the estimation of $\alpha$ of course deteriorates with decreasing recording times, but the loss of accuracy down to $T_\mathrm{obs}=250$ ms remains limited (not larger than 0.15). We therefore conclude that our machine learning strategy delivers good estimates of the value of $\alpha$ even in the case of CTRW motion. The accuracy for the estimation of the diffusion coefficient of BM motions is even better. The MAE values are around 0.70 for long $T_\mathrm{obs}$, a very good performance given that the real value is sampled uniformly at random in $(0,10]$. Here again the accuracy decreases with smaller observation times, but even with the smaller value used here, $T_\mathrm{obs}=250$ ms, the error is less than twice the error with $T_\mathrm{obs}=2$ s. 
   %Regarding the sensitivity to the beam waists, the quality of the estimation of $\alpha$ is again remarkably independent of the beam waist.     

We compared the accuracy of our method with the standard methodology of FCS, that is based on non-linear fitting of the auto-correlation function. Indeed, for BM and fBM, theoretical expressions can be derived for the decay of the auto-correlation function~\cite{hofling2013anomalous}:
    \begin{equation}
    \bar{G}_\mathrm{BM}(\tau) = \frac{1}{N} \left( 1 + \frac{4D \tau}{{\omega_{xy}}^2} \right)^{-1}  \left( 1 + \frac{4D\tau}{{\omega_z}^2} \right)^{-1/2}
    \end{equation}
and
    \begin{equation}
	\bar{G}_\mathrm{fBM}(\tau) = \frac{1}{N} \left( 1 +\left( \frac{4D \tau}{{\omega_{xy}}^2}\right)^\alpha \right)^{-1}  \left( 1 +\left( \frac{4D\tau}{{\omega_z}^2}\right)^\alpha  \right)^{-1/2}
    \end{equation}
Fitting the expression corresponding to the \textit{a priori} model of motion of the measured auto-correlation function allows one to estimate the value of the free parameters $\alpha$ and/or $D$. However, to our knowledge, such an expression is not available for CTRW, so this method cannot be used for parameter estimation in CTRW. We show in figure~\ref{fig:mae}c a comparison of the accuracy obtained using the above non-linear fits with the one obtained with our machine learning method. Both methods were applied to individual (non-averaged) auto-correlation functions like those shown in fig.~\ref{fig:F1 score}a2,b2,c2. Given the level of noise present in these recordings, it is not surprising that the estimation of $\alpha$ by standard non-linear fitting is not very good, with accuracies that are 3- to 4-times lower than our machine learning approach (Fig.~\ref{fig:maefit}a). 
%Note that to be fair, part of this limited accuracy of the fit method stems from the fact that, for CTRW, we used the expression for $\bar{G}_\mathrm{fBM}(\tau)$ above to fit the autocorrelation function, whereas this expression is not expected to be valid for CTRW. 
For the estimation of $D$, the accuracy of the non-linear fits is markedly better (Fig.~\ref{fig:mae}d). Our ML approach is still approx. 1.8-times more accurate than the standard non-linear fit method at very small $T_\mathrm{obs}$, but the accuracy values of both methods converge at long  $T_\mathrm{obs}$. Therefore, the machine-learning approach proposed in the current study demonstrates better accuracy on individual (non-averaged) synthetic FCS recordings than the standard non-linear fit methods.

    \subsection{Monitoring fast variations of the motion parameters}
    %The standard FCS methodology consists in accumulating a large number of individual measurements and averaging them in order to deal with the low SNR of individual measurements. Accumulating hundreds of measurements based on individual recordings of several seconds each thus means a total experiment duration that can several tens of minutes. It is, therefore, not feasible with standard FCS to monitor changes in the motion if their time scale is below several tens of minutes. Our results above, however, demonstrate that machine learning-based methods can be used for classification and inference of individual FCS measurements of short duration (100-200 ms). We therefore tested whether our method could be used to monitor rapid changes of the parameters of motion. 
    
\begin{figure}
    \center
        \includegraphics[scale = 1]{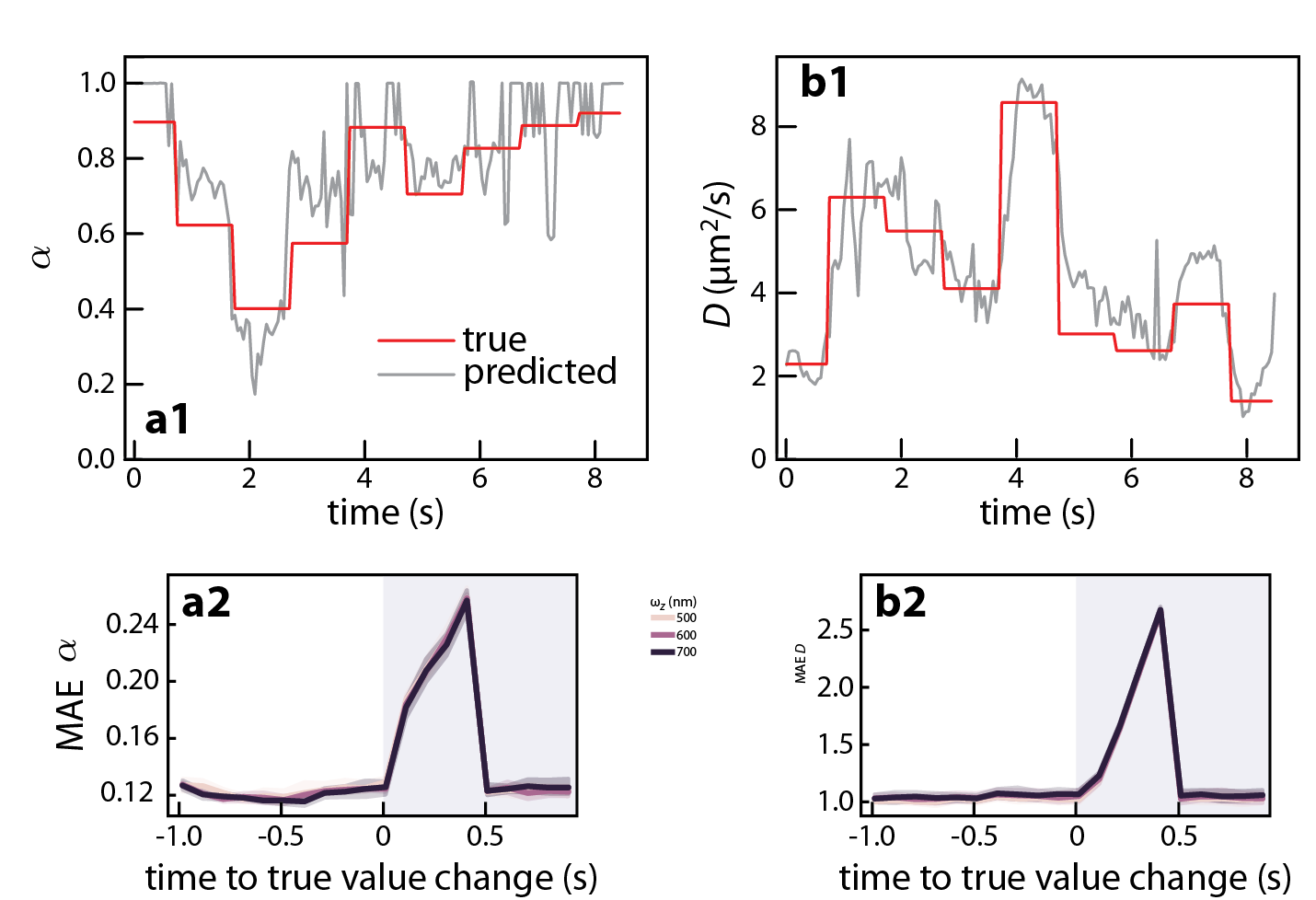}
        
        \caption{{\small\ Tracking changes of the motion parameters. 10 seconds FCS recording of CTRW (\textbf{a}) or BM (\textbf{b}) trajectories were simulated as described in sec.~\ref{mm:models}, except for the value of parameters $\alpha$ or $D$, that were not fixed but resampled from a uniform distribution every second (constant-by-parts \textit{red} curves in \textit{a1,b1}). The resulting 10-s simulated FCS recording was cut into consecutive overlapping segments of 0.5 s duration, with a 0.1 s shift and our machine-learning algorithm was applied to estimate the corresponding values of the parameters (\textit{gray} curves in \textit{a1,b1}). Estimation of the accuracy in each chunk of 1 second between two successive parameter changes is shown via the MAE of $\alpha$ (\textit{a2}) and $D$ (\textit{b2}), grouped by beam waist values $\omega_z$ as indicated in the legend. Shaded areas locate $\pm 1$ standard-deviation.}}
        \label{fig:avol}
    \end{figure}

     We then explored whether our method could be used to monitor rapid changes of the parameters of motion. To this end we used the simulation methodology presented in section~\ref{mm:algo} to generate synthetic FCS recordings of 10 s duration, where we changed the parameter of motion every second. For CTRW motion, we resampled the value of the anomalous exponent $\alpha$ every second according to an uniform distribution in $(0,1)$. For BM, we resampled the coefficient of diffusion $D$ with the same frequency, using an uniform distribution in $(0,10]$. Figure~\ref{fig:avol}a1 and b1 show examples of the resulting constant-by-part evolution of the real values of both parameters (\textit{red}). We then applied our algorithm as a sliding window of length 500 ms with a shift of 100 ms after every prediction. Figure~\ref{fig:avol}a1 shows the corresponding estimations of the anomalous exponent for the CTRW case (\textit{gray} trace). The estimation follows the changes of the true value well, with occasional delays and over estimations especially for large values of real $\alpha$ (>0.85), where our algorithm tends to classify the trajectory as BM, thus setting $\alpha$ to exactly 1. On average, however, the estimation error is large only for the first 500 ms after the parameter change, where the sliding window of the segment overlaps two true values (fig.~\ref{fig:avol}a2). Outside of these 500 ms period of overlap, the MAE converges back to the value exhibited with constant $\alpha$, i.e. around 0.13 for $T_\mathrm{obs}=0.5$ s (compare with fig.~\ref{fig:mae}a). The estimation appears slightly better for the estimation of $D$, that follows the changes of the true value quite closely (fig.~\ref{fig:avol}b1). Like for $\alpha$, the mean error on $D$ drastically increases for the first 500 ms after the change of the true value and then returns to low values (fig.~\ref{fig:avol}b2), reaching MAE values similar to those obtained with constant values of the true $D$ (fig.~\ref{fig:mae}b). 

%    We therefore conclude from these simulations that our method can indeed be applied to track changes in the motion parameters in $D \in (0,10]$ and $\alpha \in (0,1)$ of the molecules by FCS, even if the frequency of the changes is of the order of the Hz.

    % To do

    \subsection{Application to the analysis of experimental data}
    The previous series of results show that our approach provides a robust and accurate solution to motion classification and inference tasks using synthetic FCS recordings. Interpreting these results as a first validation of our method, we applied it on real experimental data. To this aim, we carried out experimental FCS measurements of calibrated 40 nm fluorescent beads in water with an increased concentration of glycerol (see section~\ref{sec:exp_data}). We applied our algorithm on these 1 second measurement as sliding window of length 500 ms with a shift of 100 ms after every prediction.

    \begin{figure}
        \includegraphics[scale=1]{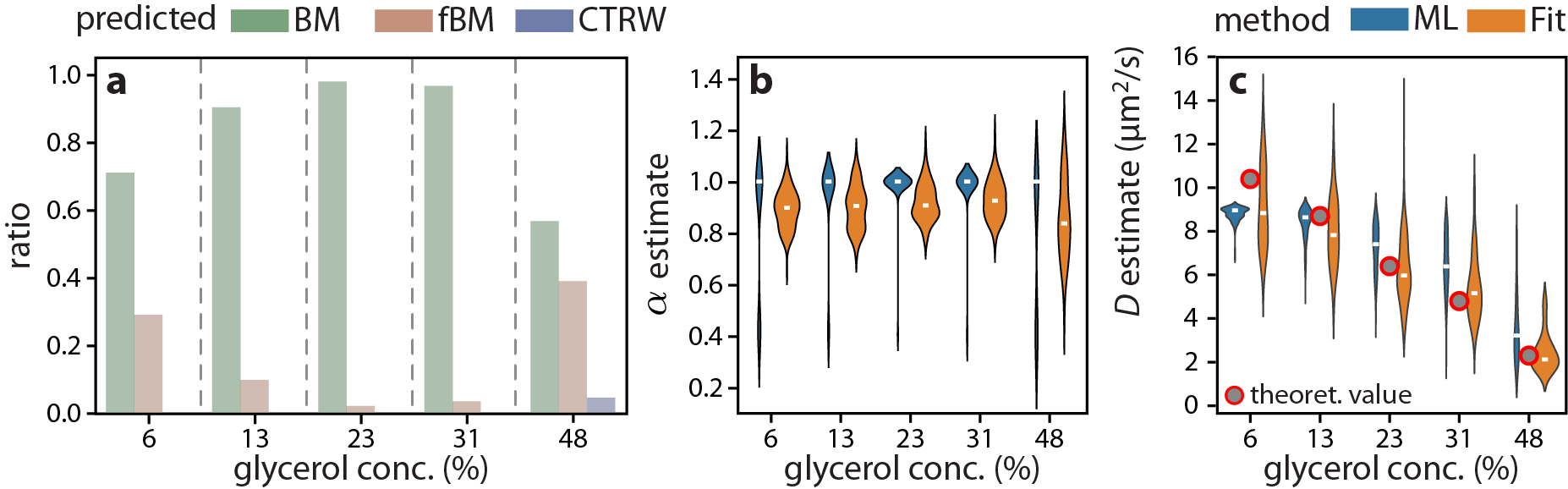}
   		 \caption{{\small Validation of our machine learning approach to experimental data. The motion of calibrated fluorescent beads in glycerol solutions of increasing concentration was monitored for 1 seconds by FCS (see section~\ref{sec:exp_data} for details). This recording was cut into 10 consecutive overlapping segments of duration 0.5 s, with a 0.1 s shift. Our machine learning approach was then used for the model classification task (\textbf{a}) and the parameter estimation task (\textbf{b-c}) on each segments. For increasing glycerol concentration, we plot in (\textit{a}) the ratio of 0.5 s segments motions that were predicted as BM (\textit{green}), fBM (\textit{brown}) or CTRW (\textit{blue}). We also show the estimations of the anomalous exponent $\alpha$ (\textit{b}) and of the diffusion coefficient $D$ for BM cases (\textit{c}) with comparison between the estimations given by our ML method and a classical non-linear fit. }}
   		 \label{fig:beads}
    \end{figure}

    Figure~\ref{fig:beads}a shows the results of the classification task with an increasing concentration of glycerol. With a small concentration of added glycerol (6\%), our algorithm classifies most of the motion segments as BM (70\%), while a minority is classified as fBM (30\%). The corresponding estimation of $\alpha$ evidences a mostly uni-modal distribution for 6\% glycerol (Figure~\ref{fig:beads}b, blue), with BM motion at $\alpha=1$. The algorithm also predicts the presence of a residual population with anomalous motion (fBM, with $\alpha$ values around 0.40). The inferred diffusion coefficient $D$ (Fig.~\ref{fig:beads}c, blue) also exhibits an unimodal distribution centered around 9 µm$^2$/s, a value that underestimates the theoretical value of 10.4 µm$^2$/s for this glycerol concentration (red-grey circles). Note that we have trained our algorithm with values of $D \in (0,10]$ µm$^2$/s, so the theoretical value of the diffusion coefficient of the beads in 6\% glycerol, 10.4 µm$^2$/s, is slightly beyond our training range. It is therefore not surprising that our estimations lack accuracy for such low glycerol concentrations. However, with increasing glycerol values, the theoretical value of $D$ is expected to decrease, and enter the training range $(0,10]$. Therefore, we expect to get better results with larger glycerol concentrations. Accordingly, the fraction of BM segments strongly increases with glycerol concentration so that the fraction of BM segments is larger than 90-95\% for 13 to 31 \% glycerol (Fig.~\ref{fig:beads}a). In this range of glycerol concentrations, the inference of $\alpha$ remains mostly concentrated around 1 (Fig.~\ref{fig:beads}b) and the distributions of $D$ exhibit medians that are close to the theoretical values (Fig.~\ref{fig:beads}c). For the largest glycerol concentration tested (e.g. 48\%), the algorithm predicts a balanced mix of mostly BM and fBM together with rare CTRW motions (less than 10\%). In addition to a majority Brownian population at $\alpha=1$, the inference of $\alpha$ again predicts an anomalous minority population centered on $\alpha=0.4$. The inference of $D$ remains very good compared to its theoretical value. Therefore our algorithm classifies the bead motions as mostly BM up to 31\% glycerol with inferred $D$ values that match their theoretical values predicted from Stokes-Einstein's law. For higher concentrations, however - here 48\% glycerol, the motions seem to become more complex, with a significant population of weakly anomalous (fBM) motion.

    In opposition to the results obtained with our method, the estimations of $\alpha$ and $D$ obtained with standard non-linear fits show much broader distributions, with medians of anomalous exponents centered around 0.8 to 0.9 (Fig.~\ref{fig:beads}b, orange). Estimations of the diffusion coefficient with this classical fitting method (Fig.~\ref{fig:beads}c, blue) appear closer to the expected theoretical values in terms of medians. However, the distributions of the estimations of $D$ are much broader than our ML estimations. Taken together, these data confirm that our ML methodology is more adapted than the standard non-linear fit for short and individual FCS measurements such as those used in these experiments, in particular because it is less biased towards slightly anomalous motions. 

\begin{figure}
    \center
        \includegraphics[scale=1.7]{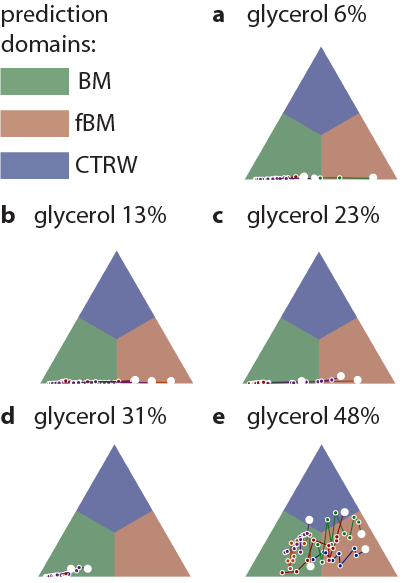}
   		 \caption{{\small Segmentation of bead motions in glycerol solutions. A ternary diagram is used to map the decision regions of the classification algorithm as a 2d representation: an FCS recording which projection falls in the green region is classified as BM by the algorithm, whereas it is classified as fBM or CTRW whenever its projections is located in the brown or blue region, respectively. The successive 0.5 s segments taken from the 1 s FCS recordings of the bead motions of fig.~\ref{fig:beads} were projected as trajectories using the ternary diagram coordinate (full lines). The initial segment is shown as a white dot. Each panel corresponds to a different concentration of glycerol: 6 (\textit{a}), 13 (\textit{b}), 23 (\textit{c}), 31 (\textit{d}) or 48 \% (\textit{e}).}}
   		 \label{fig:beadstraj} 
    \end{figure}

    We then pushed the analysis further and carried out segmentation of the FCS measurements. To this end, we projected the decision regions of our classification algorithm on a two-dimensional representation. Figure~\ref{fig:beadstraj} shows the results of this projection as a ternary diagram where the green region shows the zone where the algorithm decides that the motion is BM, whereas the brown and blue regions show where the decision is fBM or CTRW, respectively. These regions locate positions where the probability of following one model of motion is larger than the probability of following any of the other two motions. To locate the experimental FCS measurements in this 2d-plane, we projected a given experiment as a trajectory made of the classifications given by the successive sliding windows in this ternary coordinate system (full lines with full circles). With low glycerol concentrations (fig.~\ref{fig:beadstraj} a-d), most of the segments are located or at least end up in the BM domain. For some of the trajectories, the first segment or the first two segments can occasionally be found in the fBM domain, but in all cases, the trajectory quickly converges to the BM domain after this initial segment. This suggests that the minority fraction of segments classified as fBM in Fig.\ref{fig:beads}a is probably due to a lower accuracy for the classification of the very first segments in the trajectories. Inspection of the trajectories obtained with larger glycerol concentrations (48\%), confirms the results of fig.~\ref{fig:beads}a. These trajectories remain in the center of the triangle, indicating that classification is harder than the other glycerol concentrations (the difference of probabilities between two models is smaller). In addition, the trajectories are more spread out over the regions than for the other concentrations, so that a trajectory can switch classification regions several times, and not only after the first segments, as seen with 6\% glycerol. This suggests that with 48\% glycerol, the bead motions change and become more complex, in particular with the appearance of a marked heterogeneity of the motion conditions either along time or along the explored space.

\section{Discussion}

The current study is a first step to widen the applicability of Fluorescence Correlation Spectroscopy (FCS) by using machine learning for FCS recording analysis. We propose a method that is robust enough to be generic regardless of the specific technical characteristics of the setup under consideration. Depending on the laboratory or even on the specific experiment, the value of the beam waists (in $x, y$ or $z$) or the total brightness can vary. Our machine-learning algorithm has been designed to accommodate a range of values for these parameters. Figures~\ref{fig:F1 score},~\ref{fig:mae} and~\ref{fig:avol} demonstrate the performances of our algorithm over a wide range of beam waists (from 200 to 300 nm in $x,y$ and from 400 to 600 nm in $z$) on synthetic data. The limited dispersion of the resulting performance curves suggests that our method is largely independent of the exact value of the beam waists and should be applicable to a wide gamut of beam sizes.  We conclude that our machine learning algorithm should be able to accommodate many experimental setups. That being said, the algorithm cannot be expected to exhibit correct performance for technical characteristics that differ significantly from the value ranges used in the training set. 
%For instance, this would be the case for STED-FCS experiments~\cite{Sezgin2019}, where smaller values of the beam waists are reached (from 100 down-to 30 nm)~\cite{Eggeling2008}. 
In such a case, the accuracy of our approach, trained on the current parameter ranges, will likely deteriorate. This is for instance the case with our bead experiments with 6\% glycerol where the theoretical diffusion coefficient is above the range used for training the algorithm (Fig.~\ref{fig:beads}c). For these cases, our algorithm delivers a deteriorated accuracy. However, it is easy to generate a new synthetic learning set with parameter ranges that are better adapted to the specificity of the setup. We provide in parallel with the current article an open-source computer code that can be directly used to generate a new learning set, and train a new version of the algorithm on this more adapted learning set (see section~\ref{mm:code}). 
%Furthermore, approaches similar to the one used here can be exploited for the analysis of other experimental methods that are derived from FCS, such as modulation of the illumination with alternating on/off periods to model fluorescent lifetime correlation spectroscopy (FLCS)~\cite{yu2021comprehensive}.

The performance of our algorithm for the model classification and parameter inference tasks on FCS recording can be compared to the algorithms developed for the same tasks on single-particle tracking. To this end, the benchmark provided by the anomalous diffusion (AnDi) challenge is especially useful~\cite{Nature}. This collaborative open community competition has produced a fair benchmarking of the performance of more than 10 state-of-the-art algorithms on synthetic single-particle tracks (SPT). The proposed tasks included a model classification task (among 5 possible anomalous diffusion models), an inference task (anomalous diffusion exponent $\alpha$) and a segmentation task in which the model class is altered along the trajectories. Because the data used in this challenge were individual single-particle trajectories, the performance of the algorithms was quantified as a function of the most critical parameter, the length $L$ of the trajectories. It is not possible to directly use the same reference in FCS data, which do not explicitly feature trajectory length. However, since the average length of the imaged trajectories in FCS is expected to increase with the observation time $T_\mathrm{obs}$, we use $T_\mathrm{obs}$ below as a FCS proxy for $L$ in SPT. For the classification task, the best-in-class SPT algorithms exhibit $F_1$ scores ranging from 0.6 ($L=40$) to 0.9 ($L>500$) whereas for our FCS-based algorithm, the $F_1$ scores for classification varied is larger than 0.88 for $T_\mathrm{obs}>1$ second (fig.~\ref{fig:F1 score}d). Regarding the inference task, the best SPT methods provided MEA values for $\alpha$ ranging from 0.35 ($L=40$) down-to $\approx 0.10$ ($L>500$). For comparison, even if we exclude the case of incorrectly classified BM (~\ref{fig:mae}a, brown), the MEA of our FCS method for the estimation of $\alpha$ varied from 0.14 ($T_\mathrm{obs}=0.25$ s, fBM) to circa 0.11 ($T_\mathrm{obs}=2$ s, CTRW). We conclude from these comparisons that our FCS-based machine-learning approach exhibits performances that are similar to the best-of-the-class SPT algorithms of the AnDi challenge. Our methods may even be a bit better for short $T_\mathrm{obs}$ than SPT methods on short $L$. However, the limit of these comparisons is that the tasks are not entirely similar: we sampled $\alpha \in (0,1]$, compared to $[0.05,2]$ in the AnDi challenge and our set of possible motions take into account 3 models, instead of 5 models in the AnDi challenge. This differences preclude a precise one-to-one comparison, so we only retain the general conclusion that our method on FCS data yields an accuracy that favorably compares to the best-of-the-class methods for SPT data.

%In our analyses above, we did not estimate the diffusion coefficient $D$ for anomalous diffusion (fBM and CTRW). The reason for this lies in part in practicality, since it allows to keep constant the number of free parameters for all the motions considered. But more importantly, the estimation of $D$ in anomalous diffusion is usually a very difficult problem because errors on the estimation of $\alpha$ strongly deteriorate the quality of the estimation of $D$. This is probably due to the fact that the value of $\alpha$ not only impacts the value of $D$, but also its unit, which is distance$^2$/time$^\alpha$. An interesting perspective of our work is therefore to develop further methods dedicated to accurate inference of $\alpha$ and $D$ in anomalous motions. Using deep learning instead of machine learning could, for instance, be an interesting alternative. In our present case, where only the estimation of $\alpha$ was targeted, replacing our machine-learning algorithm (histogram gradient boosting) by a deep-learning approach did not improve the classification. We leave for future work the exploration of the possibility that, with a larger learning set and the inference of both $\alpha$ and $D$, deep learning might provide better predictions.    

As a validation of our method, we applied it to experimental FCS measurement of calibrated fluorescent beads in solutions with an increasing glycerol concentration. For all the studied glycerol concentrations but the largest one (48\%), our algorithm predicts that the bead motion essentially remains Brownian with a diffusion coefficient that decreases with an increase of glycerol. This is in agreement with the behavior expected from the diffusion of spherical molecules at very low Reynolds numbers in viscous fluids or from point tracers among diffusing mobile obstacles (see e.g.,~\cite{Dauty2004}). Our estimates for the diffusion coefficient $D$ also agree with the values one would expect from the Stokes-Einstein's law. 
%This is an important test for our method because one experimental FCS study of protein diffusion in aqueous solutions of a polymer as macro-molecular crowder has reported unexpected predictions of anomalous diffusion~\cite{banks2005anomalous}. Our present study cannot shed much light about these results, but we note that applying the standard nonlinear fitting procedure on our experimental bead data also concludes to weakly anomalous diffusion, in opposition to the conclusion provided by our algorithm. This suggests that the standard FCS fitting procedure may lead in some cases to an overestimation of the motion anomality. 
However, with very large glycerol concentrations (48\%), our algorithm reports a change in the bead motions, that start to depart from pure Brownian. Further work is required to confirm the signification of these observations, though, but we hope that the method introduced in the present article will be helpful to this end.

\section{Online Methods}

\subsection{Machine learning methods}
The goal of our machine learning approach is to (\textit{i}) learn to predict the class of motion $M$ of the random walkers among the set of possible motions $\mathcal{M}=\{\mathrm{BM, fBM,CTRW}\}$ (classification task), and (\textit{ii}) estimate the value of the parameters $\theta_M$ of this motion, i.e. $D$ for BM and $\alpha$ for fBM and CTRW.  

\subsubsection{Auto-correlation functions}
Our analysis starts with the collection of photon emission times, $\{\Gamma(t), t\leq T_\mathrm{obs}\}$ that constitutes the raw data of an FCS experiment (see sec.~\ref{mm:FCS}). $T_\mathrm{obs}$ is the total measurement duration. Let $E = \left(\mathcal{S},\mathcal{F},\mathbb{P}\right)$ be a probability space with sample space $\mathcal{S}$, event space $\mathcal{F}$ and probability function $\mathbb{P}$. In case of a stationary process (true for BM and a fBM), $\Gamma$ is $L^2\left([0,T_\mathrm{obs}],E\right)$, in the sense that $||\Gamma||_{L^2} = \int_0^{T_\mathrm{obs}} \left.  \mathbb{E}\left[|\Gamma(s)|^2\right]\right. \mathrm{d}s < +\infty$. In this case, $\Gamma$ admits an auto-correlation function~\cite{brockwell2009time} denoted $\left\lbrace G(\tau),\tau\in [0,{T_\mathrm{obs}}-\tau] \right\rbrace$ that depends on the auto-correlation lag $\tau$ but not on time $t$:
 \begin{equation}
    \label{eq:Goftau_base}
    G(\tau) = \frac{ \left\langle \Gamma(t)\Gamma(t+\tau)\right\rangle - \left\langle \Gamma(t)\right\rangle\left\langle \Gamma(t+\tau)\right\rangle }{\sqrt{ \left\langle \Gamma^2(t)\right\rangle \left\langle \Gamma^2(t+\tau)\right\rangle}}
    \end{equation}
where $\left\langle \cdot \right\rangle$ denotes ensemble averaging. \newline

To introduce time binning, we first define a few notations: 
\begin{itemize}
    \item Number of photons emitted between $t_a$ and $t_b$: $I[t_a,t_b] = \int_{t_a}^{t_b} \left.\Gamma(s) ds \right.$ 
    \item Bin interval: $\Delta \tau = \frac{{T_\mathrm{obs}}}{L} $ , where $L$ the length of the binned vector
    \item Binned value of $I$: $\left({I\left[i\right] }\right)_{i \in [0,L-1]} = \left({I\left[i \Delta \tau,(i+1) \Delta \tau\right] }\right)_{i \in [0,L-1]}$
\end{itemize} 

Using these notations, we estimate the ensemble-average $\left\langle \Gamma(t)\right\rangle$ of eq.~\ref{eq:Goftau_base} by its time-average $\Bar{I} = \frac{1}{L} \sum_{i=0}^{L-1} I[i]$ and its second moment $\left\langle \Gamma^2(t)\right\rangle$ by ${\Bar{I}}^2$, since $\left\langle {\Gamma^2(t)}\right\rangle = {\left\langle \Gamma(t)\right\rangle}^2$ for a Poisson process. This leads to an approximation of $G$ by its time-averaged auto-correlation estimator $\hat{G}$~\cite{hamilton2020time}:

\begin{align}
        \label{eq:Gest}
        \hat{G}(\tau) = \frac{1}{L-\tau/\Delta \tau}\sum_{i=0}^{L-1-\tau/\Delta \tau} \frac{\left. I\left[i\right] I\left[i+\tau/\Delta \tau\right] \right.-{\bar{I}}^2}{{\bar{I}}^2}
    \end{align}
    
In case $\Gamma$ is not stationary but still $L^2\left([0,{T_\mathrm{obs}}],E\right)$, i.e. for the CTRW in our case, the auto-correlation function eq.~\eqref{eq:Goftau_base} is not defined, but it is still possible to construct a partial auto-correlation function~\cite{Lambert-Lacroix1998,degerine2002evolutive,degerine2003characterization} for every associated $t \in [0,{T_\mathrm{obs}}]$, denoted as $\left\lbrace G_t(\tau),t,\tau\in [0,{T_\mathrm{obs}}-\tau] \right\rbrace$. The partial auto-correlation function of such a non-stationary process is a quantity characterizing the autocorrelation function of the stationary process associated to the non-stationary process for every $t$, defined by :
    
     \begin{equation}
    G_t (\tau) =  \frac{ \left\langle \Gamma(t)\Gamma(t+\tau)\right\rangle - \left\langle \Gamma(t)\right\rangle\left\langle \Gamma(t+\tau)\right\rangle }{\sqrt{ \left\langle \Gamma^2(t)\right\rangle \left\langle \Gamma^2(t+\tau)\right\rangle}} , 
	\end{equation}
In theory, the partial auto-correlation function of a non-stationary process can not be estimated by time averaging, but only by ensemble averaging~\cite{dix2006fluorescence}. This is not suitable in our case since we want to produce estimations for each trajectory. However, we still used the time-averaging of eq.~\ref{eq:Gest} as a feature to quantify the auto-correlation of non-stationary processes based on the \textit{ansatz} that this feature is still good enough for machine learning algorithms. This \textit{ansatz} originates from the hypothesis that the process $\Gamma$ exhibits periodicity at long times, which would mean that the mean on $t$ of its partial auto-correlation function  
\begin{equation}
	m(\tau) =\lim\limits_{T_\mathrm{obs} \to \infty}\frac{1}{{T_\mathrm{obs}}-\tau} \int_0^{{T_\mathrm{obs}}-\tau} \left.  G_s(\tau) \right. ds
\end{equation}
exists and is finite. In this case, the quantity $\hat{G}(\tau)$ from eq.~\eqref{eq:Gest} is also a good estimator for non-stationary processes.

As a final step, we normalize the feature $\hat{G}(i \Delta t)$ obtained from eq.~\eqref{eq:Gest} by dividing it by the mean of its first five elements and reduce dimensionnality by keeping only the first $K<\frac{L}{2}$ values of the sequence, using log sampling of the delay $\tau$. 

%Therefore, with $\tau_{\min}$ and $\tau_{\max}$ the min and max auto-correlation delays considered, the feature we used to quantify each FCS simulation can be written as:

%$$\Bar{G}_i  = \frac{\hat{G}\left(10^{\log_{10}(\tau_{\min})} 10^{i/K \log_{10}(\tau_{\max} / \tau_{\min})}  \right)}{\frac{1}{5} \sum_{j=1}^4 \hat{G}(j)}, \; i \in [0,K]$$.

\subsubsection{Machine learning methods}\label{mm:ML_methods}
\textbf{Learning set}. A central concern in this work is that our machine learning methods must be robust to the variety of setups used in experimental labs and, in particular, must be able to be generalized to a range of beam waists $\omega_{xy}$ and $\omega_z$. To this aim we generated a learning set comprising more than 2.5 million simulated FCS experiments of various duration and beam waists, in the following way:
\begin{itemize}
\item We first set the value of the motion parameters with uniform sampling: $\alpha \sim \mathcal{U}((0,1))$ and $D \sim \mathcal{U}((0,10])$
\item Using the algorithms described in section~\ref{mm:motions}, we then generated three sets of simulated trajectories using the sampled $\alpha$ and $D$: one with fBM motion, one with CTRW motion and one with BM motion (for BM, we set $\alpha=1$).
\item For each resulting set of trajectories, we sampled the corresponding set of photon emission times for 3 seconds, using the thinning algorithm of section~\ref{mm:FCS}. The process of photon time sampling was repeated with all possible pairs of beam waists among $\omega_{xy} \in \{200, 225,250,275,300\}$ nm and $\omega_z \in \{400,500,600\}$ nm, resulting in 15 FCS simulations per set of trajectories.
 \item In order to analyze the performance of our machine learning algorithms depending on the duration of the FCS experiment, every 3s FCS simulation described above was split into non-overlapping segments of duration $T_\mathrm{obs} \in \{0.1, 0.25, 0.5, 0.75, 1,1.25,1.5,2\}$ seconds, and every one of the 60 resulting segments was used in the learning set. With this procedure, the number of examples in the learning set was larger for short $T_\mathrm{obs}$ than longer ones (e.g., 10 times more examples with $T_\mathrm{obs}=0.1$ s compared to $T_\mathrm{obs}=1.0$ s). This allowed us to invest more learning effort on shorter observation times than larger ones. 
\item Finally, we computed the estimator of the auto-correlation $\hat{G}$ from eq.~\eqref{eq:Gest} for each of the simulation fragments above.
\end{itemize}
We repeated this process 945 times (i.e., 945 samplings of the motion parameters), yielding a learning set of 2,551,500 simulated FCS experiments in total. This learning set was then split into a test set (315 sampled parameter values, i.e. 850,500 simulations, 4.8\% of the total) and  a training set (the rest of the simulations) using uniform distribution.

\textbf{Learning algorithm}. The initial feature associated with each simulated FCS experiment is a vector of size 1,003, comprising the 1,000 log-sampled values of $\hat{G}$, plus the values of $\omega_{xy}$, $\omega_z$ and $T_\mathrm{obs}$ used for this simulation. We used these features to train a classifier $C$ with the Histogram Gradient Boosting Classifier of scikit-learn\cite{scikit-learn} (sklearn.ensemble.HistGradientBoostingClassifier) with default parameters. The classifier yields the predicted model probability for the simulation: $C(\bar{G},T_\mathrm{obs}, \omega_{xy}, \omega_z) = (\mathbb{P}_\mathrm{BM},\mathbb{P}_\mathrm{fBM},\mathbb{P}_\mathrm{CTRW})$.   

In a second phase, we trained regressors to determine $\alpha$ and $D$ (Histogram Gradient Boosting Regressor of scikit-learn with default parameters), individually for each pair of $\omega_{xy}$ and $\omega_z$ and each candidate model. For each pair of values ($\alpha$, $D$), this resulted in $5\times 3 \times 3 = 45$  classifiers, $\left(R_{\omega_{xy},\omega_{z},M}\right)$.  The input to these classifiers is also the vector of size 1,003: $(\bar{G},T_\mathrm{obs}, \omega_{xy}, \omega_z)$. For example $R_{225,600,\mathrm{fBM}}$ is trained on data with beam waist diameter of $\omega_{xy}=225$ nm, $\omega_z=600$ nm and with diffusion model fBM. These regressors are trained to predict $\alpha$ and $D$:
    \begin{equation}
    R_{\omega_{xy},\omega_z,M}(\bar{G},T_\mathrm{obs}, \omega_{xy}, \omega_z)= 
    		\begin{cases}
    			\hat{\alpha} & \text{if } M \in \{\mathrm{fBM,CTRW}\} \\
    			\hat{D}  & \text{if } M = \mathrm{BM}
    		\end{cases}
    \end{equation}
    
The final stage consolidates the classification and the regression tasks above using a last Histogram Gradient Boosting Regressor that takes into account the regression estimation for all beam waists and model classes. This final regressor $\bar{R}$ learns to predict $\alpha$ and $D$ taking as input the output of the above classifier $C$ and the outputs of the 45 corresponding regressors (vector of size 3+45+3=51): $\left(R_{\omega_{xy},\omega_z,M}\right)$:

    \begin{equation}
    \begin{split}
    \bar{R} \left(C(\bar{G},T_\mathrm{obs}, \omega_{xy}, \omega_z),\left(R_{\omega_{xy},\omega_z,M}(\bar{G},T_\mathrm{obs}, \omega_{xy}, \omega_z) \right),T_\mathrm{obs}, \omega_{xy}, \omega_z\right) \\
    		= \begin{cases}
    			\hat{\alpha} & \text{if } M \in \{\mathrm{fBM,CTRW}\} \\
    			\hat{D}  & \text{if } M = \mathrm{BM}
    		\end{cases} 
    \end{split}
    \end{equation}
    
For inference or testing, we determine the model class according to the maximal value of $\mathbb{P}_M$ estimated by the classifier $C$ and the estimation of the parameter value ($\hat{\alpha}$ or $\hat{D}$) according to the prediction of the final regressor $\bar{R}$.

\subsubsection{Code availability}\label{mm:code}
The entirety of the code used in the present article is available as an open source framework at \url{{https://gitlab.inria.fr/nquilbie/mlfcs}}. In particular, the repository offers the possibility to download the trained algorithm for use on a local computer. We also provide the code needed to generate a personalised synthetic learning set and to train the algorithm on it. The repository also proposes a simple interface based on a Jupyter notebook that allows the user to upload their own data (either as direct FCS recordings or the derived auto-correlation functions) and use our trained algorithm for the classification and inference tasks. The gitlab repository comes with a medium-size test set of synthetic trajectories, that can be used to test the performance of the algorithm. The whole training set used here (more than 2.5 million synthetic trajectories), or the experimental FCS measurements of the beads represent a considerable volume of data. The corresponding files are too large to be made available on a open access server, but they can be obtained from the authors upon request.

\subsection{Experimental data}
\label{sec:exp_data}
 To evaluate our estimation method, we tested it on experimental data. We carried out FCS measurements with calibrated fluorescent beads. FCS measurements were performed using a confocal microscope (Nikon A1R) with a 488 nm diode laser (LBX-488, Oxxius). Experiments were conducted with polystyrene nanobeads (Fluoro-Max G40, Thermo Fisher) with an average diameter of 40 nm and diluted in a water-glycerol mixture to modulate the viscosity and, consequently, the diffusion coefficient. The sample was placed in a glass bottom dish (0.16-0.19 mm, P35G-1.5-20-C MatTek) and FCS measurements were acquired using a 40x NA = 1.25 water immersion objectif (CFI Apo LWD Lambda S). The beam waists were determined as $\omega_{xy}=214$ nm and $\omega_z=522$ nm. The output signal from the sample was collected with a photon counting module (SPCM-CD, Excelitas), and time tagging was carried out by a time-correlated single photon counting module (HydraHarp 400, PicoQuant). Bead solutions were diluted to reach a concentration of $10^{11}$ particles/mL, resulting in approximately 2 individual beads on average within the focal volume.
 
Assuming that the beads in glycerol solutions are spherical objects and the flows are dominated by the viscous effect, the Reynolds numbers is very small ($Re <<1$). Then, the theoretical value of their diffusion coefficient can be estimated using the Stokes-Einstein formula $D=kT /(6 \pi \cdot \eta_0 \cdot r_\mathrm{beads})$ where  $\eta_0$ is the viscosity of the glycerol solution and $r_\mathrm{beads}$ the bead radius. We estimated the dependence of the viscosity $\eta_0$ to glycerol concentration according to Ref~\cite{cheng_formula_2008,volk_density_2018,bosart_specific_1928}. Using $r_\mathrm{beads}=20$ nm in the Stokes-Einstein formula then yields a theoretical estimate for the bead diffusion coefficient.

\newpage
\setcounter{page}{1}

\renewcommand{\theequation}{SI.\arabic{equation}}
\renewcommand{\thesection}{SI}
\renewcommand{\thesubsection}{\thesection.\arabic{section}}

\setcounter{equation}{0}
\setcounter{section}{0}
\section{Supplementary Information}

\subsection{Generation of synthetic FCS data}\label{mm:algo}
\subsubsection{Models of random motion}\label{mm:models}
\label{mm:motions}
This study focuses on models for anomalous diffusion, \textit{i.e.}, random motions for which the mean squared displacement $\left\langle r^2(t) \right\rangle$ scales non-linearly with time: 
	\begin{equation}
	\label{msd}  
	\left\langle r^2(t) \right\rangle = 2dD t^\alpha
	\end{equation}

where $r(t)$ is the position of ta random walker at time $t$, $\left\langle \cdot \right\rangle$ denotes ensemble averaging (averaging over a population of walkers at time $t$), $\alpha \in (0,1]$ is the anomalous coefficient, $D \in \mathbb{R}_+$ the diffusion coefficient and $d$ the dimension of the space (here $d=3$). The literature refers to motions with $\alpha < 1 $ as ``subdiffusive'' \textit{vs} ``superdiffusive'' for $\alpha > 1$ ($\alpha = 1$ being standard BM) \cite{METZLER20001,hofling2013anomalous,Woringer2020}. 
    
We note $W_i$ the waiting time between the ${i-1}^{th}$ and the $i^{th}$ jumps of the random walker and consider 
 	$\left(W_i \right)_{\left( i\geq 1 \right)}$ the associated i.i.d family of random variables of density $\lambda$. We associate it with the jump time  $J_i$ of the $i^{th}$  jump:
	\begin{equation}
	J_i = \sum_{n=1}^{i} \left.  W_n \right.  \text{ with } i\geq 1
	\end{equation}

Let $\Delta_i \in \mathbb{R}^d$ be the vector in space representing the $i^{th}$ displacement in space. We note $\left(\Delta_i \right)_{\left( i\geq 1 \right)}$ the corresponding family of random variables, of law ${\Delta X}_i$. The position of the particle in the $d$-dimensional-space at time $t$, $r(t)$, with initial position $r_0\in \mathbb{R}^d$ is
	\begin{equation}
	r(t) = r_0 + \sum_{i \geq 0} \left. \Delta_i \mathds{1}_{\left\lbrace J_i < t \right\rbrace}  \right.
	\end{equation}

Consider a walker located at position $x$ at time $s$, that has arrived there at time $J_i=t-s$. With these notations, the next jump of the walker will happen at time $J_{i+1} = t - s + W_{i+1}$, and its new position will be $x + \Delta_{i+1}$.

In the current study, we focus on three motion models, that we define below for the spatial dimension $d=1$:
    \begin{itemize}
    \item Brownian motion (BM) \cite{gillespie1996mathematics} is a stationary process with independent Gaussian increments: $\lambda = \delta_{dt}$, with $dt$ the simulation time step. For BM, $\left(\Delta_i \right)= \left(\mathcal{N}\left(0, \sqrt{2 D dt}\right)\right)$ is an i.i.d. Gaussian random variable family $\forall i$,     $\alpha = 1$.
    
    \item Fractional Brownian motion (fBM) \cite{mandelbrot1968fractional,mandelbrot1971fast,coeurjolly2000simulation,dieker2004simulation}, which is also a stationary Gaussian process but different from white noise due to the temporal auto-correlation of its increments: $\lambda = \delta_{dt}$, $\mathbb{E}\left[\Delta_i\Delta_j\right]
 = D\left( {|i dt|}^\alpha + {|j dt|}^\alpha + {|i dt - jdt|}^\alpha \right)$, $\alpha < 1$.  
    
    \item Continuous time random walk (CTRW) \cite{schulz2014aging,mainardi2000fractional}, which also has Gaussian distributed jumps, but which is not a stationary process if the distribution of its residence times is heavy-tailed, for instance according to a power-law:
     $\lambda(t) =\frac{\alpha}{\epsilon}  {\left(\frac{\epsilon}{\epsilon+t}\right)}^{\alpha + 1}$, $(\Delta_i) = \mathcal{N}\left(0, \sqrt{2 D dt}\right), \, \forall i$, $\alpha < 1$. We used $\epsilon = 10^{-7}$ throughout this work.
    \end{itemize}

In this study, random walks were simulated in $d=3$ space dimensions by simulating a $d=1$ independent random walk for each of the 3 dimensions. The random walks were simulated in a sphere $\Omega $ of diameter $\{\Omega_{x},\Omega_y,\Omega_z\}$ centered on $(x,y,z)=(0,0,0)$. Their initial location was uniformly distributed in $\Omega$. To keep a constant density of walkers in $\Omega$, some form of boundary condition has to be imposed at the surface of the sphere. We rejected reflective boundaries because they induce artificial correlations that strongly impact the auto-correlation signal. Instead, we used the following condition: whenever a walker leaves the sphere, we remove it from the simulation and replace it by a new walker, the initial location of which is chosen at random over the surface of the sphere.

\subsubsection{Modelling of FCS measurements}
\label{mm:FCS}
We simulated an FCS illumination volume centered at $(0,0,0)$, the center of the spherical domain $\Omega$ in which the random walks occur. The point spread function (PSF) of the microscope is modelled as a 3d Gaussian with beam waists $\omega_{i} << \Omega_{i},\;\forall i\in\{x,y,z\}$ \cite{krichevsky2002fluorescence}. In agreement with the experimental situation we considered identical beam waists in the $x$ and $y$ directions, i.e. $\omega_{x}=\omega_{y}\coloneqq\omega_{xy}$. The illumination intensity $\Phi$ is thus given by 
    \begin{equation}
    \Phi(x,y,z) =\Phi_0 e^{-2 \left({\frac{x^2+y^2}{{\omega_{xy}}^2}+\frac{z^2}{{\omega_{z}}^2}}\right)},
    \end{equation}
    where $\Phi_0$ controls the illumination intensity.
    
The probability that a particle located at $(x,y,z)$ emits a photon is modelled as a Poisson process with a rate proportional to the value of the illumination at this position~\cite{horton2010development,krichevsky2002fluorescence}. Since the particle location changes according to the random walk, we model photon emission by a single walking particle as a non-homogeneous Poisson process~\cite{ng2023non}, with time-dependent rate $\mu(t) = \Phi\left(r(t)\right)$.
	
If $\gamma = \left\lbrace \gamma(t), t \geq 0 \right\rbrace$ is the process characterizing the times of photon emission by a single molecule, one has 
	\begin{equation}
	\left( \gamma(t) \right)_{t \geq 0} = \partial \mathcal{P}\left( \left( \mu(t) \right)_{t \geq 0} \right)
	\end{equation}
	
where $\partial \mathcal{P}$ is the process of the jump times of a Poisson process, i.e. if $\left( \Theta_i \right)_{i\geq 0}$ are the jump times of $\mathcal{P}$, then $\partial \mathcal{P} = \sum_{i \geq 0} \delta_{\Theta_i}$ 	. Now, to retrieve $\Gamma = \left\lbrace \Gamma(t) , t\geq 0 \right\rbrace$, the counting process characterizing the emission times of photons from an FCS experiment with $N$ molecules, we sum the $N$ processes characterizing each molecule $\Gamma(t) = \sum_{n=1}^N \gamma_n(t)$. By the additive property of Poisson processes ($\partial\mathcal{P}(\sigma) +\partial \mathcal{P}(\nu) = \partial\mathcal{P}(\sigma+\nu)$~\cite{kingman1992poisson}), the intensity of the system can be modelled as the sum of the intensities of the $N$ processes:
	\begin{equation}
	\bar{\mu}(t) =  \sum_{n=0}^N \left.\mu_n(t)\right.  
	\text{ and } 
	\Gamma = \partial\mathcal{P}\left(  \left( \bar{\mu} (t) \right)_{t \geq 0} \right)
	\end{equation}

     We simulate the process of photon emission by all the $N$ molecules, $\Gamma$, by thinning~\cite{lewis1979simulation}. We suppose that its rate $\bar{\mu}(t)$ is bounded for $t \in [0,T_\mathrm{obs}]$ by its maximal value  $||\bar{\mu}||_{\infty} < +\infty$. We first sample the photon emission times that would be expected from a Poisson process with constant (homogeneous) rate $||\bar{\mu}||_{\infty}$: $\tilde{\Gamma} = \partial \mathcal{P} \left( \left( ||\bar{\mu}||_{\infty} \right)_{t \geq 0} \right)$. We refer to those as candidate emission times $\tilde{T_i}$: 
     \begin{equation}
         \tilde{\Gamma} = \sum_{i \geq 0} \left. \delta_{\tilde{T_i}} \right. \sim \partial\mathcal{P}\left(  ||\bar{\mu}||_{\infty} \right)
     \end{equation} 	
     
     We then reject some of the candidate emission times $\tilde{T_i}$ to adapt them to $\bar{\mu}(t)$: we associate with every candidate emission time $\tilde{T_i}$, a uniformly-distributed random variable $U_i \sim \mathcal{U}\left(0,||\bar{\mu}||_{\infty}\right)$ and reject every $\tilde{T_i}$ for which $U_i > \bar{\mu}(\tilde{T_i})$. The emission times that were not rejected thus define the photon emission times of our initial process: 
     \begin{equation}
         \Gamma = \sum_{i \geq 0} \left. \delta_{\tilde{T_i}}\cdot \mathds{1}_{ U_i \leq \bar{\mu}(\tilde{T_i})} \right. \sim \partial\mathcal{P}\left(  \left( \bar{\mu} (t) \right)_{t \geq 0} \right)
     \end{equation} 	
     
The output of the simulation is the resulting collection of the times the $N$ random walkers emitted photons, $\left( T_i \right)_{i\geq 0}$. Note that the above process is currently in continuous time, but it will be binned later during pre-processing.

\subsubsection{FCS simulation parameters}
For the simulations of the current paper, we used the following parameter values:
\begin{itemize}
    \item Size of the spatial domain $\mathcal{D}$  ($\mu$m): 
 $(\mathcal{D}_x,\mathcal{D}_y,\mathcal{D}_z)= (1.05,1.05,2.4)$
    \item Time step $dt= 1 \, \mu$s 
    \item Beam waist ($\mu$m): $\omega_{xy} \in \{0.200,0.225,0.250,0.275,0.300\}$, $\omega_z \in \{0.500, 0.600, 0.700\}$. 
    \item Mean number of random walkers in the illuminated volume $v(=4/3  \pi  \omega_{xy}^2 \omega_z)$: $n=5$ 
    \item Maximal illumination $\Phi_0=6\times10^4$
\end{itemize}

\bibliography{ref}
\bibliographystyle{ieeetr}

\end{document}